\documentclass[a4paper,11pt]{article}
\usepackage{pos}
\usepackage{hyperref}

\usepackage{amsmath}
\numberwithin{equation}{section} 
\usepackage{gensymb}
\usepackage{mathtools}

\usepackage{caption,subcaption, array} 
\usepackage{float} 

\usepackage{braket} 
\usepackage{mhchem} 
\usepackage{siunitx} 
\usepackage[compat=1.1.0]{tikz-feynman} 

\newcommand{\mrm}[1]{\mathrm{#1}}
\newcommand{\bb}[1]{\mathbb{#1}}
\newcommand{\op}[1]{\hat{#1}}
\newcommand{\vecop}[1]{\vec{#1}}

\renewcommand{\d}[1]{\mathop{}\!\mathrm{d}{#1}\mathop{}\!}

\title{\(\ce{p}\Omega\) femtoscopy using baryon-baryon effective potentials}

\author*[a]{M. Piquer i Méndez}
\author[b]{A. Parreño García}
\author[b]{J. Torres-Rincon}

\affiliation[a]{Facultat de Física,\\
	Universitat de Barcelona, Martí i Franquès 1, 08028 Barcelona, Spain}

\affiliation[b]{Dept. de Física Quàntica i Astrofísica and Institut de Ciències del Cosmos, Facultat de
Física,\\
Universitat de Barcelona, Martí i Franquès 1, 08028 Barcelona, Spain}

\emailAdd{mpiquer13@gmail.com}
\emailAdd{assum@fqa.ub.edu}
\emailAdd{torres@fqa.ub.edu}

\abstract{We have generated an updated version of the \(\ce{p}\Omega\) potential for low-energy interactions based on an effective field theory approach at leading order. This potential, together with other potentials based either on different parametrizations or lattice QCD calculations, have been used to solve the Schrödinger equation numerically, obtaining the scattering wave functions for different values of the relative momentum. Using these wave functions, we have computed the \(\ce{p}\Omega\) femtoscopic correlation functions, comparing the results with those published by the ALICE collaboration.
}

\FullConference{%
  QNP2024 - The 10th International Conference on Quarks and Nuclear Physics\\
  8-12 July, 2024\\
  Barcelona, Spain
}

\begin{document}
\maketitle

\section{Introduction}

A usual choice to study hadronic interactions in the non-perturbative regime of quantum chromodynamics (QCD) are  effective field theories (EFTs)~\cite{Weinberg}. In these approaches, quarks and gluons are replaced by hadrons as fundamental degrees of freedom, describing the interactions among them below certain energy scale. An effective Lagrangian is written in terms of all possible operators which respect the symmetries of the problem, leading to a systematic expansion in the momenta of the external particles. To study the dynamics of light baryons and mesons, one can exploit the baryon chiral perturbation theory (\(\chi\)PT) to derive an effective potential for the baryon-baryon interaction. In this work we will focus on the proton-$\Omega$ baryon (\(\ce{p}\Omega\)) pair.

While experimental information of the hadronic interaction is primarily obtained through scattering experiments, these are really challenging when some of the involved particles are unstable, as it is the case of hyperons, such as the $\Omega$ baryon. Femtoscopy studies~\cite{FMV} have been very helpful in obtaining relevant information in these cases, where some of the features of the interaction are mapped in the shape of the pair correlation function $C(\vec{k})$, where $\vec{k}$ is the relative momentum in the center-of-mass frame. This function can be expressed in terms of the scattering wave function \(\psi (\vec{r},\vec{k}) \) and the so-called source function \(S(\vec{r})\) through the Koonin-Pratt formula~\cite{FMV},
\begin{equation} \label{Koonin-Pratt}
	C(\vec{k}) = \int_{\bb{R}^3} S(\vec{r}) |\psi(\vec{r},\vec{k})|^2 \d{\vec{r}} {\rm .}
\end{equation}
In our approach to the femtoscopy study of the \(\ce{p}\Omega\) interaction, we will solve the Schrödinger equation to obtain the scattering wave functions for appropriate potentials---derived from the EFT or from numerical calculations, such as lattice QCD (LQCD)---and we will compare it to the experimental \(\ce{p}\Omega\), obtained by the ALICE collaboration using high-energy \(\ce{p}\ce{p}\) collisions at the Large Hadron Collider~\cite{ALICE}. This work is a summary of \cite{TFM}, whose results were presented at this conference. All data plotted in it can be found in \href{https://github.com/Marc-PiM/Masters_thesis_plot_data}{https://github.com/Marc-PiM/Masters\_thesis\_plot\_data} (2024).


\section{Effective potential, scattering wave-function and correlation functions}

\begin{figure}[H]
	\centering
	\begin{subfigure}[c]{0.3\textwidth}
		\centering
		\resizebox{!}{2 cm}{%
        \feynmandiagram {
			f1 [particle=\(\Omega^-\)] -- [opacity = 0.0] fmid -- [opacity = 0.0] f2 [particle=\(\ce{p}\)],
			m1 -- [opacity = 0.0] ct [dot] -- [opacity = 0.0] m2,
			i1 [particle=\(\Omega^-\)] -- [opacity = 0.0] imid -- [opacity = 0.0] i2 [particle=\(\ce{p}\)],
			i1 -- [fermion] ct -- [fermion] f1,
			i2 -- [fermion] ct -- [fermion] f2,
		};
        }
		\subcaption{Contact diagram.}
	\end{subfigure}
	\begin{subfigure}[c]{0.3\textwidth}
		\centering
		\resizebox{!}{2 cm}{%
        \feynmandiagram {
			f1 [particle=\(\Omega^{-}\)] -- [opacity = 0.0] f2 [particle=\(\ce{p}\)],
			a -- [charged scalar, edge label=\(\eta\)] b,
			i1 [particle=\(\Omega^{-}\)] -- [opacity = 0.0] i2 [particle=\(\ce{p}\)],
			i1 -- [fermion] a -- [fermion] f1,
			i2 -- [fermion] b -- [fermion] f2,
		};
        }
		\caption{\(\eta\) meson exchange diagram.}
	\end{subfigure}
	\begin{subfigure}[c]{0.3\textwidth}
		\centering
		\resizebox{!}{2 cm}{%
        \feynmandiagram {
			f1 [particle=\(\Omega^{-}\)] -- [opacity = 0.0] f2 [particle=\(\ce{p}\)],
			a -- [double, with arrow = 0.5, edge label=\(\sigma\)] b,
			i1 [particle=\(\Omega^{-}\)] -- [opacity = 0.0] i2 [particle=\(\ce{p}\)],
			i1 -- [fermion] a -- [fermion] f1,
			i2 -- [fermion] b -- [fermion] f2,
		};
        }
		\caption{\(\sigma\) meson exchange diagram.}
	\end{subfigure}
	
	\caption{Feynman diagrams contributing to the \(\ce{p}\Omega\) scattering at LO in the EFT.}
 \label{EFT-diagrams}
\end{figure}
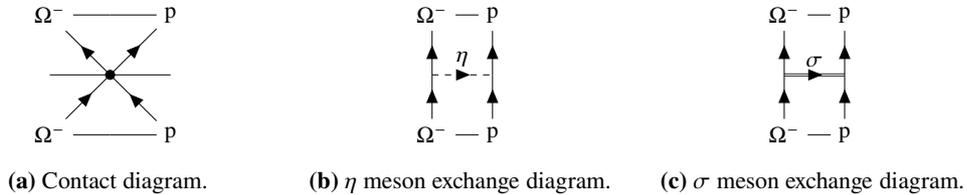

After writing down the baryon $\chi$PT Lagrangian we apply the Weinberg power-counting~\cite{Weinberg} to organize the diagrams of the \(\ce{p}\Omega\) interaction. Keeping only the leading order (LO) terms, we end up with the diagrams shown in Fig.~\ref{EFT-diagrams}: (a) contact terms without derivative couplings, (b) and (c) one-meson-exchange diagrams, restricted to the pseudoscalar $\eta$ and scalar $\sigma$ mesons. These are the lightest exchanged mesons allowed by charge, isospin and strangeness conservation at each vertex.
The spin operators allowed in the contact term are the unit operator, ${\hat 1}$, and the product of the spins of the baryons, $\vec{\mathbb{S}}\cdot\vec{\sigma}$. We regularize the contact terms using a Gaussian {\it ansatz}, introducing the regulator $\delta = \sqrt{2}/m_{\omega}$, where the ultraviolet scale is set by the $\omega$ meson mass, which is the first vector meson absent in our calculation. Finally, at each vertex we incorporate a form factor (FF),  \(F_{\Lambda} (\vec{q}) = \Lambda^2 / (\Lambda^2 + {\vec{q}\,}^2)\), which depends on the cut-off parameter $\Lambda$, to be fixed.

\begin{figure}[H]
	\centering
	\begin{subfigure}[b]{0.40\textwidth}
		\centering
		\includegraphics[width = \textwidth]{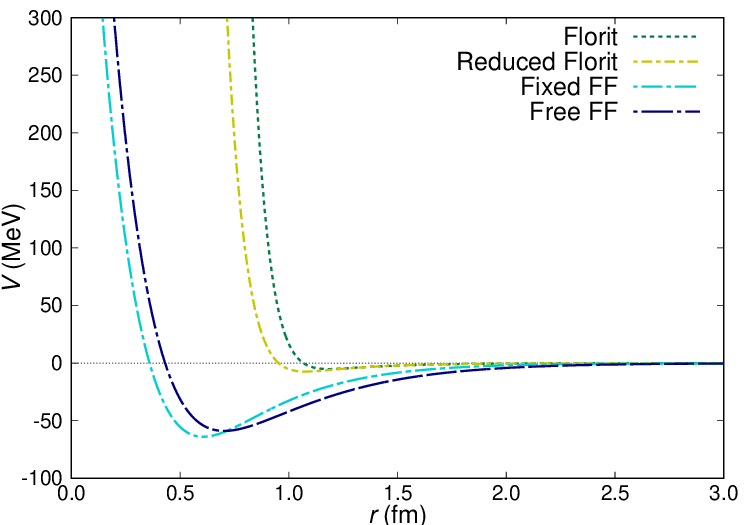}
		\caption{\(\prescript{3}{}{S}_1\) channel.}
	\end{subfigure} \qquad \begin{subfigure}[b]{0.40\textwidth}
	\centering
	\includegraphics[width = \textwidth]{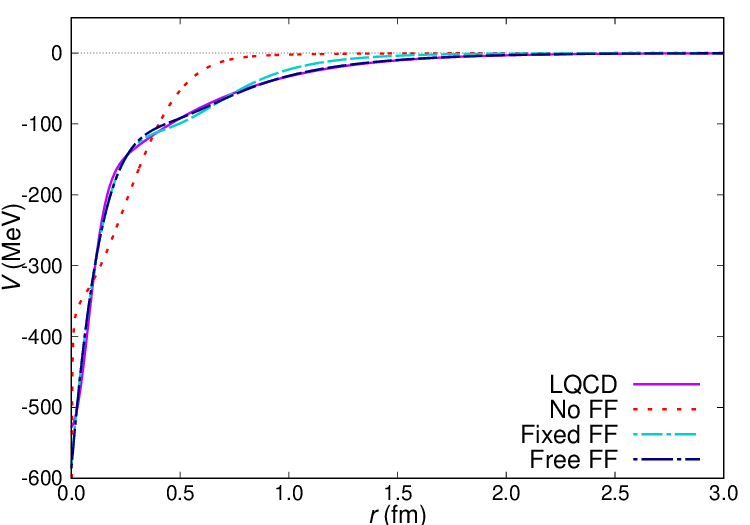}
	\caption{\(\prescript{5}{}{S}_2\) channel, fit parametrizations.}
	\end{subfigure}
	
	\caption{\(\ce{p}\Omega\) strong potentials (without Coulomb forces), for the different parametrizations used in Tab.~\ref{Tab:Fits}. For comparison we include the \(\prescript{5}{}{S}_2\) potential taken from LQCD~\cite{HALQCD}.}\label{fig:pots}
\end{figure}

Performing a Fourier transformation to coordinate space, we obtain the elastic p$\Omega$ potentials,
\begin{align}
\op{V}_{\mrm{ct}}^{\mrm{el}} &= \left(C_0^0 + C_0^1 \vecop{\mathbb{S}}\cdot\vecop{\sigma}\right) \frac{e^{-(r/\delta)^2}}{\pi^{3/2} \delta^3} \ , \nonumber\\
\op{V}_{\eta}^{\mrm{el}} &= C_{\ce{p}\eta\bar{\ce{p}}} C_{\Omega\eta\bar{\Omega}} \frac{{m_{\eta}}^2}{3} \left(\frac{{\Lambda_{\eta}}^2}{{\Lambda_\eta}^2 - {m_{\eta}}^2}\right)^2 \left[\frac{e^{-m_{\eta}r}}{4\pi r} - \frac{e^{-\Lambda_{\eta}r}}{4\pi r} + \frac{({m_\eta}^2 - {\Lambda_\eta}^2) \Lambda_\eta}{8 \pi {m_{\eta}}^2} e^{-\Lambda_{\eta} r}\right] \vecop{\mathbb{S}}\cdot\vecop{\sigma} \ , \nonumber \\
\op{V}_{\sigma}^{\mrm{el}} &= -C_{\ce{p}\sigma\bar{\ce{p}}} C_{\Omega\sigma\bar{\Omega}} \left(\frac{{\Lambda_{\sigma}}^2}{{\Lambda_\sigma}^2 - {m_{\sigma}}^2}\right)^2 \left[\frac{e^{-m_{\sigma}r}}{4\pi r} - \frac{e^{-\Lambda_{\sigma}r}}{4\pi r} + \frac{{m_\sigma}^2 - {\Lambda_\sigma}^2}{8 \pi {\Lambda_{\sigma}}} e^{-\Lambda_{\sigma} r}\right] \ ,
\end{align}
where $\vecop{\mathbb{S}}\cdot\vecop{\sigma}=\{ -5/2, 3/2 \}$ for the $^3S_1$ and $^5S_2$ channels, respectively; the couplings \(C_{\ce{p}\eta\bar{\ce{p}}} \approx 2.17\cdot 10^{-3}\:\si{MeV}^{-1}\) and \(C_{\Omega\eta\bar{\Omega}} \approx 2.36\cdot 10^{-2}\:\si{MeV}^{-1}\) are obtained from the baryon $\chi$PT~\cite{Martí}, and \(C_{\ce{p}\sigma\bar{\ce{p}}} \approx 8.71\) is estimated from the Walecka-Serot model~\cite{Walecka}. 
The remaining parameters of the model, \(C_0^0 \), \(C_0^1\), \(C_{\Omega\sigma\bar{\Omega}}\), \(\Lambda_{\eta}\) and \(\Lambda_{\sigma}\), are fitted under different approaches summarized in Table~\ref{Tab:Fits}.
\begin{table}[H]
	\centering
	\begin{tabular}{r||c|c|c|c|c|}
		& \(C_0^0 (\si{MeV^{-2}})\) & \(C_0^1 (\si{MeV^{-2}})\) & \(C_{\Omega\sigma\bar{\Omega}}\) & \(\Lambda_{\eta} (\si{MeV})\) & \(\Lambda_{\sigma} (\si{MeV})\) \\ \hline\hline
		Florit & $5.00 \cdot 10^{-3}$ & $1.00 \cdot 10^{-3}$ & \multicolumn{3}{c}{} \\ \cline{1-3}
		Reduced Florit & $1.25 \cdot 10^{-3}$ & $0.25 \cdot 10^{-3}$ & \multicolumn{3}{c}{} \\ \cline{1-4}
		No FF & $-8.06(6)\cdot 10^{-6}$ & $-1.61(2) \cdot 10^{-6}$ & $0.8858(2)$ & \multicolumn{2}{c}{} \\ \hline
		Fixed FF & $1.133(5)\cdot 10^{-5}$ & $2.267(9) \cdot 10^{-6}$ & $1.882(8)$ & $900$ & $1200$ \\ \hline
		Free FF & $6.59(6)\cdot 10^{-6}$ & $1.32(2) \cdot 10^{-6}$ & $2.58(3)$ & $917.4(7)$ & $622(6)$ \\ \hline
	\end{tabular}
	\caption{Parameter sets used in this work for the different approaches (left column). The blank cells correspond to constants not appearing in the corresponding model.} \label{Tab:Fits}
\end{table}
The ``Florit'' and ``Reduced Florit'' choices~\cite{Martí} use no \(\sigma\) exchange nor form factors, and are based on a simple scaling down of the couplings in the strangeness \(S=-1,-2\) sectors~\cite{Martí}. The remaining 3 choices fit the \(\prescript{5}{}{S}_2\) potential given in the LQCD results of Ref.~\cite{HALQCD}: ``No FF'' uses no form factors at all, ``Fixed FF'' keeps fixed the values of $\Lambda_\eta,\Lambda_\sigma$ and fits the remaining parameters, and ``Free FF'' leaves free all parameters in the fit. The resulting potentials are shown in Fig.~\ref{fig:pots}.

From the results in Fig.~\ref{fig:pots} we disregard the ``No FF'' fit, since it does not compare well with the LQCD potential. Our best fit corresponds to the ``Free FF'' one. After solving the Schr\"odinger equation, this potential presents bound states in both the \(\prescript{5}{}{S}_2\) and \(\prescript{3}{}{S}_1\) elastic channels with binding energies of \(E = -2.13\:\si{MeV}\) and \(E = -0.48\: \si{MeV}\), respectively.

\begin{figure}[H]
	\centering
	\begin{subfigure}[b]{0.40\textwidth}
		\centering
		\includegraphics[width = \textwidth]{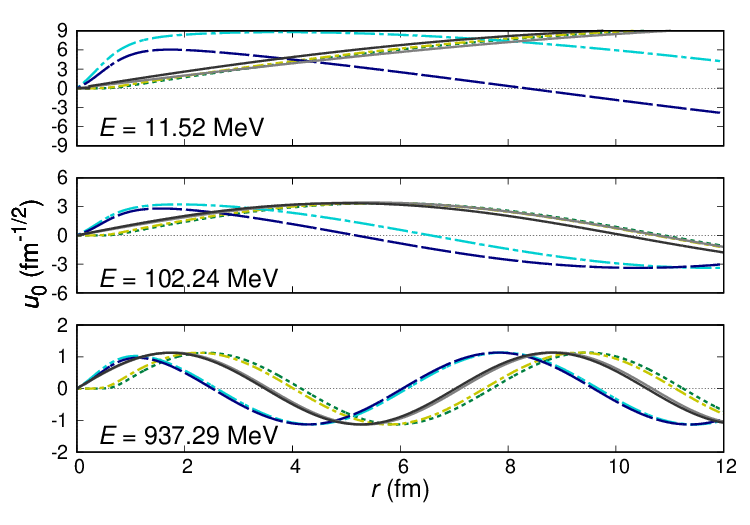}
		\caption{\(\prescript{3}{}{S}_1\) channel.}
	\end{subfigure}
	\begin{subfigure}[b]{0.40\textwidth}
	\centering
	\includegraphics[width = \textwidth]{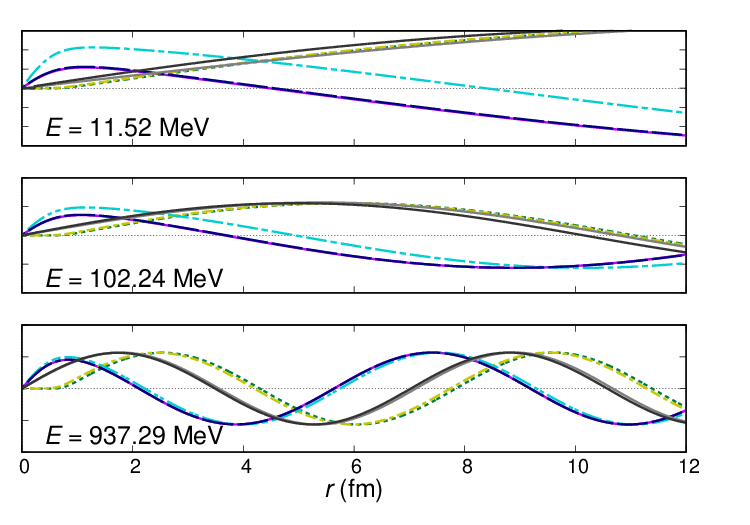}
	\caption{\(\prescript{5}{}{S}_2\) channel.}
	\end{subfigure} \\
    \begin{subfigure}[b]{\textwidth}
	\centering
    \vspace*{0.2 cm}
	\includegraphics[width = 0.18\textwidth, clip = true, trim = 100 18 180 226]{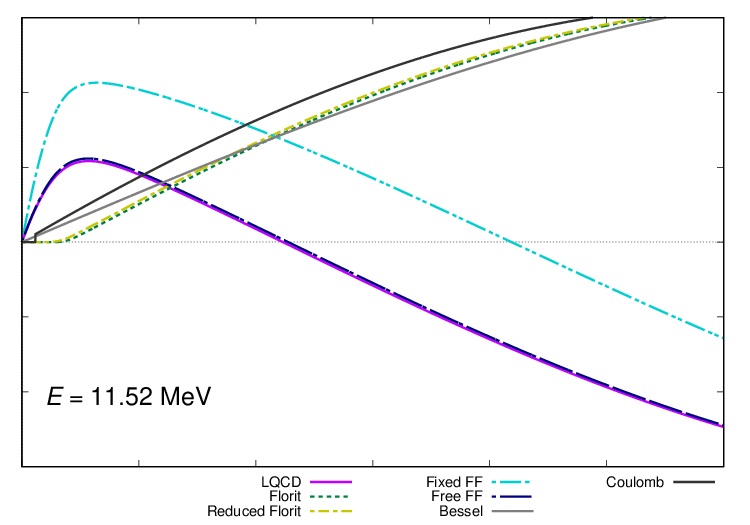} 
	\includegraphics[width = 0.18\textwidth, clip = true, trim = 100 3 180 241]{Llegenda.eps} 
    \includegraphics[width = 0.18\textwidth, clip = true, trim = 187 10 93 234]{Llegenda.eps} 
    \includegraphics[width = 0.18\textwidth, clip = true, trim = 187 2 93 242]{Llegenda.eps} 
    \\ \vspace*{-0.25 cm}
    \includegraphics[width = 0.18\textwidth, clip = true, trim = 100 10.5 180 233.5]{Llegenda.eps} 
    \includegraphics[width = 0.18\textwidth, clip = true, trim = 187 18 93 226]{Llegenda.eps} 
    \includegraphics[width = 0.18\textwidth, clip = true, trim = 274 18 6 226]{Llegenda.eps} 
    \hspace{0.18\textwidth}
	\end{subfigure}
	\caption{\(\ce{p}\Omega\) $L=0$ wave functions from the different parametrizations and the LQCD potential of Ref.~\cite{HALQCD}, together with the corresponding Bessel and Coulomb wave functions, for comparison.} \label{fig:wavefuncs}
\end{figure}

Using the parametrizations for the  \(\ce{p}\Omega\) potential given in Tab.~\ref{Tab:Fits}, we obtain the $s-$wave scattering states by numerically solving the Schr\"odinger equation. The results are summarized in Fig.~\ref{fig:wavefuncs}, where we observe great similarity between the LQCD and our best fit's, ``Free FF'', wave functions. 

In order to compute the femtoscopy correlation functions, we assume that the strong interaction only affects the $L=0$ partial wave. In this way, the Koonin-Pratt formula (\ref{Koonin-Pratt}) reduces to
\begin{equation} \label{eq:CoulKP}
		C(\vec{k}) = \int_{\bb{R}^3} S(\vec{r}) |\Phi^{\textrm{C}} (\vec{r},\vec{k})|^2 \d{\vec{r}}  + \ 4\pi \int_0^\infty  S(r)
  [|u_0(r,k)|^2 - |F_0(k,r)/k|^2] \d{r} \ , 
\end{equation}
where $\Phi^{\textrm{C}} (\vec{r},\vec{k})$ is the complete Coulomb wave function, and $F_0(k,r)$ the regular Coulomb function with $L=0$. The Fortran 90 implementation is described in detail in Ref.~\cite{TFM}.

\begin{figure}[H]
	\centering
	\includegraphics[width = 0.5\textwidth]{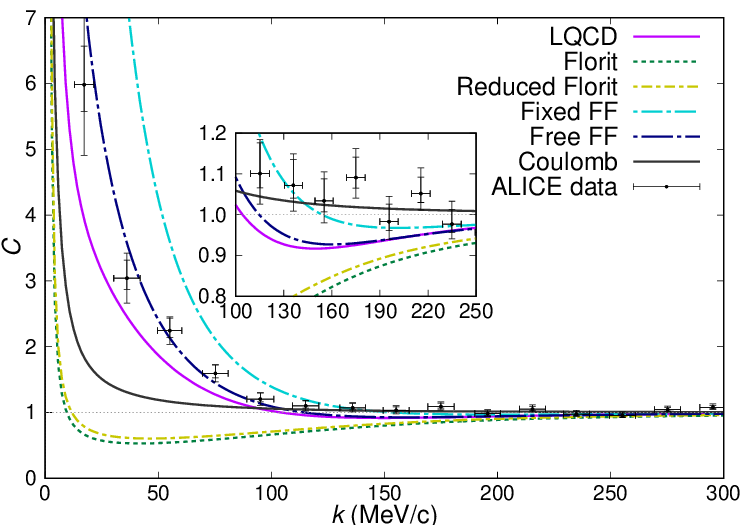}
	\caption{Correlation functions for the considered \(\ce{p}\Omega\) potentials, together with the ALICE data from Ref.~\cite{ALICE}.}\label{fig:Ck}
\end{figure}

In Fig.~\ref{fig:Ck} we show the spin-averaged correlation functions for the potentials considered, and compare with the ALICE experimental data extracted in~\cite{ALICE}. We use a Gaussian source with radius \(r_0 = 0.95\:\si{fm}\). Our best fit, the ``Free FF'' parametrization, is able to describe the experimental data, getting a slight overestimation in the low-momentum region, and a underestimation around $k=150\:\si{MeV}$ due to the effect of the bound states, not present in the experimental data.

\section{Conclusions and Outlook}

Using the ``Free FF'' fit of the \(\prescript{5}{}{S}_2\) \(\ce{p}\Omega\) potential we have also predicted the \(\prescript{3}{}{S}_1\) potential, which was absent in~\cite{HALQCD}. As for the \(\ce{p}\Omega\) correlation function, the ``Free FF'' fit slightly improves over the results from~\cite{HALQCD} when compared with ALICE data~\cite{ALICE}. Our result overestimates the data in the low-momentum region (as opposed to the results in~\cite{ALICE} using the LQCD potential) since we added the contribution of the \(\prescript{3}{}{S}_1\) channel. This fact, together with additional higher partial waves, leaves room for possible negative contributions of the inelastic channels, as opposed to the conclusions in~\cite{ALICE}. Future work will concentrate on including higher partial waves, but also the inelastic coupled channels, to test this claim. Using the same EFT we can also study the femtoscopy correlation functions of other baryon-baryon systems, like the \(\ce{p}\Xi^{-}\) or \(\Omega\Omega\) systems. 

\acknowledgments

This work has been supported by the project number
CEX2019-000918-M (Unidad de Excelencia ``Mar\'ia de Maeztu''), PID2020-118758GB-I00 and PID2023-147112NB-C21, financed by the Spanish MCIN/ AEI/10.13039/501100011033/.
MP thanks the Institut de Ciencies del Cosmos for financial support during the master's degree and his advisors AP and JT for their help.


\begin{thebibliography}{99}

\bibitem{Weinberg} S. Weinberg, \textit{Nuclear forces from chiral lagrangians}, \href{https://doi.org/10.1016/0370-2693(90)90938-3}{\textit{Phys. Lett. B} \textbf{251}: 288-292 (1990)}; S. Weinberg, \textit{Effective chiral lagrangians for nucleon-pion interactions and nuclear forces} \href{https://doi.org/10.1016/0550-3213(91)90231-L}{\textit{Nucl. Phys. B} \textbf{363}: 3-18 (1991)}.

\bibitem{FMV} L. Fabbietti, V. Mantovani Sarti and O. Vázquez Doce, \textit{Study of the Strong Interaction Among Hadrons with Correlations at the LHC}, \href{https://doi.org/10.1146/annurev-nucl-102419-034438}{\textit{Annu. Rev. Nucl. Part. Sci.} \textbf{71}: 377-402 (2021)}.
	
\bibitem{ALICE} ALICE collaboration, \textit{Unveiling the strong interaction among hadrons at the LHC}, \href{https://doi.org/10.1038/s41586-020-3001-6}{\textit{Nature} \textbf{588}: 232-238 (2020)}.

\bibitem{TFM} M. Piquer i Méndez, \textit{Addressing the \(N\Omega\) femtoscopy correlation function using baryon-baryon effective potentials} (master's thesis, Universitat de Barcelona, 2024).

	
\bibitem{Martí} M. Florit Gual, \textit{Strong baryon-baryon interaction in the strangeness \(-3\) sector} (master's thesis, Universitat de Barcelona, 2014).
	
	
\bibitem{Walecka} B. D. Serot and J. D. Walecka, \textit{The Relativistic Nuclear Many Body Problem}, \textit{Adv. Nucl. Phys.} \textbf{16}: 1-327 (1986).

\bibitem{HALQCD} T. Iritani, S. Aoki et al., \textit{\(N\Omega\) dibaryon from lattice QCD near the physical point}, \href{https://doi.org/10.1016/j.physletb.2019.03.050}{\textit{Phys. Lett. B} \textbf{792}: 284-289 (2019)}.
	
	
    
	
	

	
\end{thebibliography}
\end{document}